\documentclass[a4paper,traditabstract,letter]{aa}

\usepackage{graphicx}
\usepackage{txfonts}
\usepackage{natbib}
\usepackage{longtable,lscape}
\usepackage[footnotesize]{caption}   

\begin{document}

\title{Using spectral flux ratios to standardize SN~Ia luminosities}


\authorrunning{Bailey et al. (Nearby Supernova Factory)}

\author{
    The Nearby Supernova Factory: \\
    S.~Bailey\inst{1},       
    G.~Aldering\inst{2},     
    P.~Antilogus\inst{1},    
    C.~Aragon\inst{2},       
    C.~Baltay\inst{3},       
    S.~Bongard\inst{1},      
    C.~Buton\inst{4},        
    M.~Childress\inst{2,5},  
    N.~Chotard\inst{4},      
    Y.~Copin\inst{4},        
    E.~Gangler\inst{4},      
    S.~Loken\inst{2},        
    P.~Nugent\inst{2},       
    R.~Pain\inst{1},         
    E.~Pecontal\inst{6},     
    R.~Pereira\inst{4},      
    S.~Perlmutter\inst{2,5}, 
    D.~Rabinowitz\inst{3},   
    G.~Rigaudier\inst{6},    
    K.~Runge\inst{2},        
    R.~Scalzo\inst{3},       
    G.~Smadja\inst{4},       
    H.~Swift\inst{2,5},      
    C.~Tao\inst{7},          
    R.~C.~Thomas\inst{2,8},  
    and C.~Wu\inst{1}        
}

\institute{Laboratoire de Physique Nucl\'eaire et des Hautes Energies, CNRS/IN2P3;
Universit\'es Paris VI et Paris VII,
4 place Jussieu, Tour~43, 75252 Paris Cedex 05
\and
Physics Division, Lawrence Berkeley National
Laboratory, 1 Cyclotron Road, Berkeley, CA 94720
\and
Department of Physics, Yale University, New Haven, CT 06250-8121
\and
Universit\'e de Lyon, F-69622 France;
Universit\'e de Lyon 1;
CNRS/IN2P3, Institut de Physique Nucl\'eaire de Lyon
\and
Department of Physics, University of California Berkeley, Berkeley, CA 94720
\and
Universit\'e de Lyon, F-69622 France;
Universit\'e de Lyon 1;
CRAL, Observatoire de Lyon, F-69230 Saint Genis Laval;
CNRS, UMR 5574;
ENS de Lyon
\and
CPPM, 163 av. Luminy, 13288 Marseille Cedex 09
\and
Luis W. Alvarez Fellow
}

\begin{abstract}
{
We present a new method to standardize Type~Ia supernova (SN~Ia)
luminosities to $\la 0.13$~magnitudes using flux ratios from a single
flux-calibrated spectrum per SN.
Using Nearby Supernova Factory spectrophotomery of 58 SNe~Ia,
we performed an unbiased search for flux ratios that correlate with
SN~Ia luminosity.  After developing the method and selecting the best
ratios from a training sample, we verified the results on a
separate validation sample and with data from the literature.
We identified multiple flux ratios whose correlations with luminosity are
stronger than those of light curve shape and color, previously identified
spectral feature ratios, or equivalent width measurements.
In particular, the flux ratio
$\mathcal{R}_{642/443} = F(642~{\rm nm}) / F(443~{\rm nm})$
has a correlation of 0.95 with SN~Ia absolute magnitudes.
Using this single ratio as a correction factor produces a Hubble diagram
with a residual scatter standard deviation of $0.125 \pm 0.011$ mag,
compared with $0.161 \pm 0.015$ mag when fit with the SALT2
light curve shape and color parameters $x_1$ and $c$.
The ratio $\mathcal{R}_{642/443}$ is an effective correction
factor for both extrinsic dust reddening and instrinsic variations
such as those of SN~1991T-like and SN~1999aa-like SNe.
When combined with broad-band color measurements, spectral flux ratios
can standardize SN~Ia magnitudes to $\sim 0.12$ mag.
These are the first spectral metrics that give robust improvements over the
standard normalization methods based upon light curve shape and color,
and they provide among the lowest scatter Hubble diagrams ever published.
}
\end{abstract}


\date{Submitted 2 March 2009; Accepted 4 May 2009}

\maketitle

\section{Introduction}

The ability to standardize the absolute luminosity of Type Ia supernovae (SNe~Ia)
makes them powerful cosmological probes through their use as luminosity
distance indicators.  Uncorrected observations of SN~Ia absolute magnitudes
have an RMS scatter of $\sim$40\%; corrections for light curve
shape and color normalize them to 15\% to 20\%, {\it i.e.}, luminosity
distances of SNe~Ia can be measured to 7-10\% accuracy using current
standardization techniques \citep{Phillips93, Prieto2006, Guy_SALT2, Jha_MLCS}.
Improved standardization would reduce
both statistical and systematic errors, since the remaining dispersion 
represents an upper limit to possible bias from uncorrected physical effects
that could arise from variations in host galaxy dust properties,
progenitor properties such as mass, metallicity, and age, and explosion
physics such as geometry.

Spectral indicators such as feature ratios
\citep[{\it e.g.},][]{Nugent_RSi, Bongard_RSiS, Foley_UV} 
and pseudo-equivalent widths \citep[{\it e.g.},][]{Hachinger_EW, Bronder_EWSiII}
have been proposed to enhance
peak SN~Ia normalization from light curve shape and color,
or even as alternative approaches altogether which use
a single night's observations to correct observed magnitudes
\citep[{\it e.g.},][]{SnapshotDistances}.
While these methods showed promise for what could be acheived with spectra,
none of them is robustly competitive with light curve shape
and color corrections alone.

In this work, we take a new unbiased approach to the problem of
finding optimal spectral flux ratios for standardizing SN~Ia peak
brightness.
Flux ratios best correlated with absolute magnitudes are detected in a
training set and cross-validated on a separate subsample.  Hubble
diagram fits are compared using standard lightcurve color and
shape corrections,
using only flux ratios as standardization parameters, 
or using flux ratios combined 
with light curve color measurements.
We find several new robust spectral indicators that
outperform light curve shape and color corrections.

\section{Data and analysis}

This analysis uses spectrophotometry of 58 of the supernovae
obtained by the Nearby Supernova Factory
(SNfactory) collaboration using its SuperNova Integral Field Spectrograph
\citep[SNIFS,][]{SNF_SPIE} on the University of Hawaii
2.2-meter telescope on Mauna Kea.
This subset 
consists of SNfactory SNe~Ia for which final host-galaxy
follow-up has been
completed and which pass minimal cuts on the quality of the light curve fit.
Targets were additionally required to have a spectrum within
$\pm 2.5$ restframe days of $B$-band maximum light;
if multiple spectra were available, the one nearest maximum light was used.
Hubble residuals, colors, and light curve shape parameters were not a factor
in selecting this subsample.
These SNe are evenly distributed in the redshift range $0.02 < z < 0.09$,
which allows relative distance measurements with minimal uncertainties
from peculiar velocities.
The methods were developed on a training subset of 28 SNe before cross
checking the results on a separate validation sample of 30 SNe to ensure
that the final results are not simply overfitting statistical fluctuations
of the dataset.

The subsets have consistent distributions of
light curve width and color, 
redshifts, and Hubble residuals;
these are consistent with other published SNe~Ia
used for cosmology measurements and training lightcurve fitters.
The host galaxies include spirals, ellipticals, and irregulars,
and the host properties span an order of magnitude in
progenitor age, metallicity, and inferred amount of dust extinction. 
This is comparable to the estimated
evolution of the universal mean of these
quantities to $z \sim 1$  
\citep[{\it e.g.},][]{RiessLivio2006, KowalskiUnion2008}
thus the results from this dataset are
also likely to work well for high-$z$ SNe~Ia.

The spectra were dereddened to correct for Milky Way dust \citep{CCM, SFD},
deredshifted, and
rebinned in $c \Delta\lambda/\lambda \sim 2\,000$ km/s bins
from 350 to 850~nm (23 to 56~\AA\ bins) to equally sample
spectral features in the physically relevant velocity space
while still over-sampling SN~Ia spectral features.
The exact bin sizes used were not rigorously optimized for this analysis,
though cross checks were performed to ensure that the results are not
sensitive to the exact binning chosen.

Photometry was synthesized from the multi-epoch flux-calibrated spectra
in box filters corresponding to approximately $B$, $V$, and $R$.
Light curves were fit using SALT2 \citep{Guy_SALT2} to estimate their
restframe $B$-band peak magnitudes $m_B$ and calculate their uncorrected
$B$-band Hubble
diagram residuals (or equivalently, their uncorrected absolute magnitudes).
Alternatively, $m_B$ was synthesized directly from the spectra,
independent of an SN light curve model.  

\begin{figure}[t]
\centering
\includegraphics[width=0.8\columnwidth]{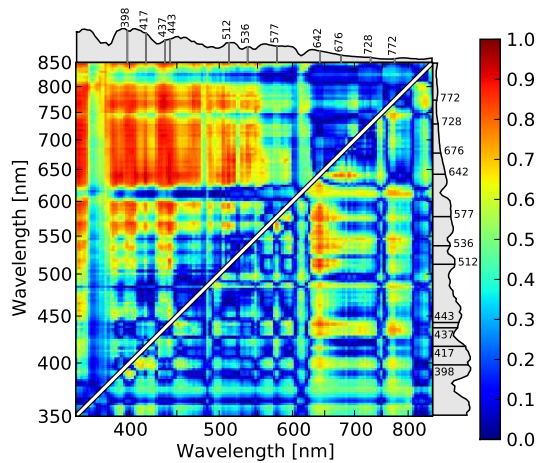}
\caption{Absolute Pearson correlations of flux ratios $\mathcal{R}_{y/x} = F_y/F_x$ with
SN~Ia absolute magnitudes.
The upper-left triangle shows the correlations with uncorrected absolute magnitudes.
For illustration, 
the lower-right triangle shows the correlations of
color-corrected ratios $\mathcal{R}^c$ with color-corrected magnitudes.
A median SN~Ia spectrum is included on each axis
for reference along with the wavelengths
used in the flux ratios of Table~\ref{tab:results}.
\label{fig:2dcorr}
}
\end{figure}

The correlation of absolute magnitudes with
each possible binned flux ratio was calculated for the training sample as
shown in Fig.~\ref{fig:2dcorr}.  
The upper-left triangle
of the color density plot shows the absolute Pearson correlation $\rho$ of
the flux ratios $\mathcal{R}_{y/x} = F_y/F_x$ with the SN~Ia absolute
magnitudes; the bottom-right triangle shows the color corrected
correlations as described below.
Using $\log(\mathcal{R})$ produced similar
though slightly worse results.

Spearman rank order correlations were also calculated
in order to be less sensitive to outliers.
Although the Spearman correlation is a more robust statistic,
the Pearson correlation is a better predictor of the effects on a 
Hubble diagram $\chi^2$ fit since outliers affect the Pearson correlation
and $\chi^2$ fits in similar ways.
We selected the best ratios using the simple average of the Pearson
and Spearman correlations; by considering both types of correlations,
we select relatively strong correlations while avoiding
those which are primarily due to outliers.  We also checked the
robustness of selected peaks by comparing the results from other near-max
spectra in the training sample which were not the spectra closest to
maximum light.

The statistical significance of prospective correlations in the training
set was assessed using Monte Carlo permutation testing before the
validation set was unblinded.
For each random trial, Hubble diagram residuals were randomized and the
ratio with the highest correlation was retained.  We find that
${\mathcal P}(|\rho| > 0.9) < 10^{-5}$, compared with multiple peaks above
0.9 in Fig.~\ref{fig:2dcorr}.

\begin{figure}[t]
\centering
\includegraphics[width=0.80\columnwidth]{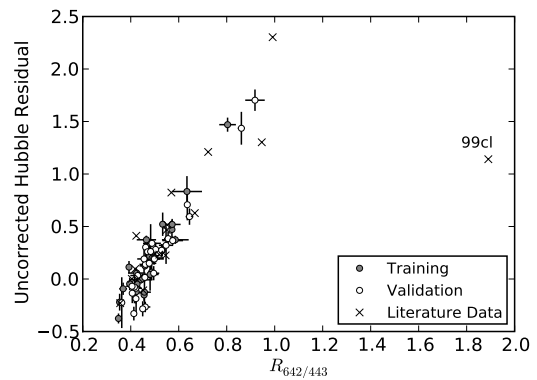}
\caption{
Hubble diagram residuals {\it vs.} $\mathcal{R}_{642/443}$ for the training set 
(filled circles), validation set (open circles),
and literature data ($\times$ symbols).
\label{fig:R_642_443}
}
\end{figure}

After selecting the five best flux ratios from the training sample, we unblinded
the results for those ratios for the validation sample.
Figure~\ref{fig:R_642_443} shows the results for the highest ranked ratio,
$\mathcal{R}_{642/443}$.  It plots the uncorrected Hubble diagram residuals $\Delta\mu_B$
{\it vs.} $\mathcal{R}_{642/443}$ for the training set of
supernovae (filled circles) and validation set (open circles).
The correlation is very strong (0.95), consistent between the two samples,
and is not dominated by just a few supernovae.
For comparison, the correlation of Hubble residuals with the combination of the
SALT2 light curve shape and color parameters $x_1$ and $c$ is 0.92.

\begin{table*}
\caption{
Correlations of flux ratios $\mathcal{R}$ with SN~Ia absolute magnitudes
and standard deviations of Hubble diagram fits.
}
\label{tab:results}
\begin{center}
\begin{tabular}{ccccrcccc}
\hline
\hline
Correction & \multicolumn{3}{c}{Correlation with Absolute Magnitude} & & \multicolumn{3}{c}{Hubble Diagram Residual Scatter} \\
Parameter(s)    & Training & Validation & Combined & $\gamma$~~~~~ & Training & Validation        & Combined & $\sigma_{\rm core}$ \\
\hline
$\mathcal{R}_{642/443}$  & 0.94 & 0.96 & 0.95 &  $3.5 \pm 0.2$  & $0.130 \pm 0.018$ & $0.134 \pm 0.018$ & $0.128 \pm 0.012$ & 0.108 \\
$\mathcal{R}_{642/417}$  & 0.95 & 0.91 & 0.91 &  $4.9 \pm 0.2$  & $0.114 \pm 0.016$ & $0.185 \pm 0.025$ & $0.166 \pm 0.016$ & 0.162 \\
$\mathcal{R}_{772/437}$  & 0.92 & 0.94 & 0.93 &  $7.3 \pm 0.3$  & $0.142 \pm 0.020$ & $0.160 \pm 0.021$ & $0.152 \pm 0.014$ & 0.125 \\
$\mathcal{R}_{642/512}$  & 0.90 & 0.95 & 0.93 &  $4.7 \pm 0.3$  & $0.162 \pm 0.022$ & $0.146 \pm 0.020$ & $0.154 \pm 0.015$ & 0.152 \\
$\mathcal{R}_{728/398}$  & 0.90 & 0.93 & 0.91 &  $7.9 \pm 0.3$  & $0.162 \pm 0.022$ & $0.168 \pm 0.022$ & $0.172 \pm 0.016$ & 0.138 \\
\hline
$c, \mathcal{R}^c_{642/519}$  & 0.96 & 0.96 & 0.96 &  $3.5 \pm 0.3$  & $0.106 \pm 0.015$ & $0.129 \pm 0.018$ & $0.119 \pm 0.011$ & 0.128 \\
$c, \mathcal{R}^c_{577/642}$  & 0.95 & 0.95 & 0.95 & $-1.4 \pm 0.1$  & $0.115 \pm 0.016$ & $0.150 \pm 0.020$ & $0.135 \pm 0.013$ & 0.126 \\
$c, \mathcal{R}^c_{642/536}$  & 0.95 & 0.96 & 0.95 &  $2.3 \pm 0.2$  & $0.116 \pm 0.016$ & $0.134 \pm 0.018$ & $0.125 \pm 0.012$ & 0.126 \\
$c, \mathcal{R}^c_{676/642}$  & 0.94 & 0.93 & 0.93 & $-4.2 \pm 0.5$  & $0.131 \pm 0.019$ & $0.178 \pm 0.024$ & $0.157 \pm 0.015$ & 0.163 \\
$c, \mathcal{R}^c_{642/443}$  & 0.95 & 0.96 & 0.96 &  $3.2 \pm 0.3$  & $0.121 \pm 0.017$ & $0.125 \pm 0.017$ & $0.119 \pm 0.011$ & 0.104 \\
\hline
$c$, $x_1$         & 0.91  & 0.93  & 0.92 & ...~~~~~  & $0.154 \pm 0.022$ & $0.171 \pm 0.023$ & $0.161 \pm 0.015$ & 0.156 \\
\hline
\end{tabular}
\newline
{\it Notes}: $\gamma$ is a fit parameter in the distance modulus
$\mu_B = (m_B - M^\prime) + \gamma \mathcal{R}$,
and $\sigma_{\rm core} = 1.4826\times {\rm median}(|\Delta\mu_B - {\rm median}(\Delta\mu_B)|)$.
\end{center}
\end{table*}

\begin{figure}[t]
\centering
\includegraphics[width=0.8\columnwidth]{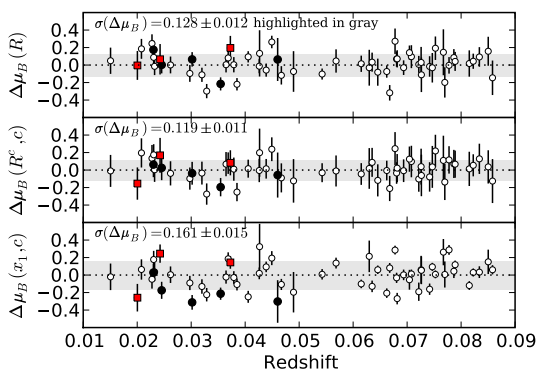}
\caption{
Hubble diagram residuals when fit with
the single parameter $\mathcal{R}_{642/443}$ (top),
with color and dereddened ratio $\mathcal{R}^c_{642/443}$ (middle),
and with the SALT2 
parameters $x_1$ and $c$ (bottom).
Gray bands highlight $\pm 1\sigma$ scatter.
Intrinsically bright SN~1991T-like and SN~1999aa-like SNe are highlighted in black;
highly dust-extincted SNe are highlighted with red squares.
\label{fig:HubbleDiagram}
}
\end{figure}

We fit a Hubble diagram of the 58 SNe using flux ratios $\mathcal{R}$ with
a distance modulus of $\mu_B = (m_B - M^\prime) + \gamma \mathcal{R}$,
where $m_B$ is the restframe $B$-band magnitude from
either SALT2 (without any color or $x_1$ corrections)
or directly synthesized from the spectrophotometry,
$M^\prime$ absorbs offsets from the $\mathcal{R}$ intercept
and the mean SN~Ia absolute magnitude,
and $\gamma$ is solved in the
fit along with $M^\prime$ to minimize the $\chi^2$ of the residuals.
The result for $\mathcal{R}_{642/443}$ when using SALT2 $m_B$
is shown in Fig.~\ref{fig:HubbleDiagram} (top).
The standard deviation of the residual scatter is
$\sigma(\Delta\mu_B) = 0.128 \pm 0.012$ mag when using
only $\mathcal{R}_{642/443}$ to standardize the peak magnitude $m_B$, compared with
$0.161 \pm 0.015$ mag when using $\mu_B = (m_B - M) + \alpha x_1 - \beta c$
instead (Fig.~\ref{fig:HubbleDiagram} bottom).
In this case, $\alpha$ and $\beta$ are allowed to vary in the fit.

Synthesizing $m_B$ directly from the flux-calibrated near-max spectra
provides a peak magnitude estimate independent of a light curve model. 
In this case $\sigma(\Delta \mu_B) = 0.125 \pm 0.011$ mag
when measuring $m_B$ and $\mathcal{R}_{642/443}$ from the same spectra.
Note that this standardization is achieved
using only a {\it single} flux calibrated spectrum per SN;
the light curve fit is used only to determine the phase.
This demonstrates the standardization power of a single spectrum and
that these flux ratios are not merely correcting for some
anomaly of the SALT2 light curve fit. 

In Fig.~\ref{fig:HubbleDiagram},
intrinsically bright SN~1991T-like and SN~1999aa-like SNe are highlighted
in black; highly extincted SNe are highlighted with red squares.
$\mathcal{R}_{642/443}$ provides an effective correction for both of these extremes
for which 
$x_1$ and $c$ are less effective.

We also repeated this analysis fitting a combination of color and flux ratios.
In this case, we dereddened the spectra using the SALT2 color model so that
the resulting spectra have $E(B-V) = 0$; this removes color from both
extrinsic dust reddening as well as possible intrinsic color
without trying to distinuish between the two effects.
The remaining spectral variations, however, should be predominantly due
to small-scale intrinsic effects and not extrinsic dust reddening.
We denote flux ratios from these decolored spectra as $\mathcal{R}^c$.
Their correlation with color-corrected Hubble diagram residuals is shown
in the lower-right triangle of Fig.~\ref{fig:2dcorr};
the final selection was based upon the combined fit of $c$ and $\mathcal{R}^c$,
not just the correlations in Fig.~\ref{fig:2dcorr}. 
Several combinations of $c$ and $\mathcal{R}^c$ have correlations
with absolute magnitude of $\sim0.95$ and result in
$\sigma(\Delta\mu_B) \sim0.12$ mag.
Figure~\ref{fig:HubbleDiagram} (middle)
shows the results for 
$\mathcal{R}^c_{642/443}$ with $\sigma(\Delta\mu_B) = 0.119 \pm 0.011$ mag.
Adding the light curve shape parameter $x_1$ to the fits does not
improve the results for the single-$\mathcal{R}$ fits or the
$c, \mathcal{R}^c$ fits.

The results are summarized in Table~\ref{tab:results}.
Table~\ref{tab:data} in the online version lists the supernovae used
with their redshifts, $x_1$, $c$, $\Delta\mu_B$, and
$\mathcal{R}^{(c)}$ values.
For all selected ratios, a single flux ratio results in
smaller or competitive Hubble diagram residual scatter compared to
what is acheived by standard corrections for light curve shape and color
($x_1$ and $c$);
the combination of a color-corrected ratio $\mathcal{R}^c$ and $c$ consistently
outperforms $x_1$ and $c$.
The dispersion of the core is also improved; flux ratios are
not merely correcting a few outliers.
The largest discrepancy between the training and validation results
is for $\mathcal{R}_{641/417}$, but even in this case, the validation set has a
residual scatter of only $0.185 \pm 0.025$ mag.
The fits are error weighted but the standard deviation is not, which
allows some cases where the combined fit has a scatter either
less than or greater than the scatter of both of the individual fits.

\section{Discussion}

The 130 to 330~nm wavelength baseline of the flux ratios means that
they are effectively color measurements using very narrow band filters,
thus they can correct for extrinsic dust
extinction in a manner similar to standard $E(B-V)$ corrections.
In fact, $\mathcal{R}_{642/443}$ has
a correlation of 0.83 with $E(B-V)$ 
and four of the five SNe with the largest $\mathcal{R}_{642/443}$ values
have host galaxy ISM features in their observed spectra,
indicating extrinsic dust extinction.

However, $\mathcal{R}_{642/443}$ also appears sensitive to intrinsic SNe variations that are
correlated with absolute magnitude.
There are other dim SNe in this sample with large $\mathcal{R}_{642/443}$ and
no spectral evidence for host extinction,
and $\mathcal{R}_{642/443}$ is able to standardize
absolute magnitudes better than color alone:
using $c$ only, the
Hubble residuals have a standard deviation of $0.215 \pm 0.020$ mag.
Additionally, the characteristic size of the correlation peaks in
Fig.~\ref{fig:2dcorr} are typical of the widths of SN features,
indicating a sensitivity to instrinsic variations of the supernovae.

PHOENIX calculations based upon the W7 explosion model
\citep{Bongard_PHOENIX} show a strong correlation of $\mathcal{R}_{642/443}$ with the absolute
$B$-band magnitude, driven by large intrinsic variations of \ion{Fe}{II}
and \ion{Fe}{III} line blends around 443~nm.
The region redward of 640~nm has
relatively little intrinsic variation in these models
and serves as a normalization anchor for the ratio.
The $\mathcal{R}_{642/443}$ correlation also appears in a new survey of 2D
SEDONA \citep{Kasen_SEDONA} 
models spanning nickel mass 0.4-1.0 $M_\odot$ with 30
different viewing angles per model; these generally reproduce the
known Phillips brighter-broader correlation.  The models exhibit a
similar correlation between $\mathcal{R}_{642/443}$ and absolute magnitude as we found,
but the correlation is even stronger than the theoretically modeled
Phillips relation (D.~Kasen, private communication).
More modeling work is needed to understand the underlying physics of
our results.

It may seem surprising that a single parameter could 
correct for the completely unrelated effects of SN~Ia intrinsic variability
and extrinsic dust reddening.
But note that the slopes for a dust correction and any intrinsic correction
will vary with different wavelength pairs selected for the flux ratio.
Some wavelength pairs will align
the slopes more than others, thus reinforcing the individual corrections and
allowing for two physically distinct phenomena to be corrected with
a single parameter.
The final best choice will also depend upon
the strength of the intrinsic correlations for that ratio and how
accurately it can be measured.

We checked our results for $\mathcal{R}_{642/443}$ using spectra within 2.5 days of
maximum from  \citet{MathesonSpec} after rejecting 3 exhibiting
strong H-$\alpha$ indicative of host galaxy contamination.
Uncertainties on $\mathcal{R}_{642/443}$ were
assigned based on the internal dispersion for the 12 SNe~Ia having
multiple suitable spectra; otherwise we used uncertainties
derived by \citet{MathesonSpec} from their comparison with photometry.
Corresponding 
$B$-band peak magnitudes for 16 SNe~Ia from \citet{Jha_LC}
and \citet{HickenLC} were put on a consistent scale and standardized
using $\mathcal{R}_{642/443}$, including photometric and redshift errors due to
peculiar velocities.
With the exception of SN~1999cl, we find good agreement with our results
with a consistent slope $\gamma$ and
strong correlation $\rho=0.93$  (see Fig.~\ref{fig:R_642_443}).
The scatter is
$\sigma(\Delta\mu_B)=0.25$~mag (0.22 for an error weighted $\sigma$), but 
the fit has $\chi^2_{\nu} = 0.9$, indicating that the dispersion
is consistent with the measurement errors for these data
and is not a contradiction of our low dispersion result.
The environment of SN~1999cl is known to be unusual,
having time-varying \ion{Na}{I~D} \citep{Blondin_99cl}
and ``highly nonstandard dust'' 
\citep{Krisciunas_99cl_dust}.
It is also a large outlier (1.5 mag) when corrected by $x_1$ and $c$.
This exception is consistent with the inability of any single 
parameter to simultaneously correct intrinsic SN variations and more than 
one type of dust behavior.

Application of $\mathcal{R}_{642/443}$ requires flux calibrated spectra near maximum
with accurate host galaxy subtraction;
spectral slope errors should be kept below a few percent.
To keep the noise contribution to the Hubble residuals under 0.1~mag,
$S/N \ga 25$ in 2\,000 km/s bins is sufficient.
For distant SNe~Ia, where $S/N$ may be a limiting constraint,
the spectra may be rebinned to a lower resolution with only
a small degradation of the results.
For instance, rebinning to 10\,000~km/s (still centered on 642 and 443~nm)
gives $\rho=0.94$ and   
$\sigma(\Delta\mu_B)=0.145$~mag.
Away from peak brightness the correlations weaken and the
standardization slope $\gamma$ changes, but
the $5(1+z)$~day window used here is adequate for scheduling spectroscopy
based on a pre-maximum lightcurve in any band.
Although currently available high redshift SN~Ia spectra are insufficient
to apply this method, future spectrophotometric instruments
in space or
on large ground-based telescopes could achieve these requirements
and use this method to build an SN~Ia Hubble diagram
with the potential for significantly improved statistical
power and systematics control.

\section{Conclusions}

We have performed a search for spectral flux ratios that correlate with
observed variability in SN~Ia
absolute magnitudes using separate training and validation datasets
totaling 58 homogeneously observed SNe with $0.02 < z < 0.09$.
The results reveal
single parameter corrections that achieve results 
better than those from light curve shape and color parameters, even when only
a single spectrum is used to measure and standardize the peak magnitude.
The combination of a flux ratio and broad-band color provides
even better results, significantly outperforming standard corrections
for light curve shape and color.
The best single flux ratio $\mathcal{R}_{642/443}$ results in a Hubble diagram 
residual scatter of $0.125 \pm 0.011$ mag; 
the best combination of color plus a ratio provides $0.119 \pm 0.011$ mag.

These results highlight the power of spectrophotometric SN~Ia observations.
Compared to lightcurve shape and color corrections,
our flux ratio standardization represents a factor of 1.8 
improvement in the statistical weight per SN,
and a 35\% improvement 
in the upper bound on the potential bias from uncorrected SN~Ia variability.
These results work
well for both extrinsic dust extinction and intrinsic varations such as
SN~1991T-like and SN~1999aa-like SNe,
and they effectively correct SNe~Ia over a range of
light curve shapes and colors, host galaxy properties, and intrinsic subtypes,
offering the promise that this method
will be similarly powerful for future high-$z$ SN~Ia observations.


\begin{acknowledgements}

We are grateful to the technical and scientific staff of the University
of Hawaii 2.2-meter telescope,
Palomar Observatory, Lick Observatory,
and the High Performance Research and Education Network (HPWREN)
for their assistance in obtaining these data.
We also thank Julien Guy and David Rubin for assistance with light curve fits.
This work was supported in France by CNRS/IN2P3, CNRS/INSU, CNRS/PNC,
and used the resources of 
the IN2P3 computer center.
This work was also supported in part by the Director,
Office of Science, Office of High Energy and Nuclear Physics, of the
U.S. Department of Energy (DOE) under Contract Nos. DE-FG02-92ER40704, 
DE-AC02-05CH11231,      
and DE-FG02-06ER06-04;  
by the Director, Office of Science, Office of Advanced Scientific
Computing Research, of the U.S. DOE under Contract No.
DE-AC02-05CH11231; 
by a grant from the Gordon \& Betty Moore Foundation;
by National Science Foundation
Grant Nos. AST-0407297 (QUEST),
AST-0606772 (CfA Supernova Archive),
and 0087344 \& 0426879 (HPWREN);
by a Henri Chretien International Research Grant
administrated by the American Astronomical Society;
and the France-Berkeley Fund.

\end{acknowledgements}

\onllongtabL{2}{
\begin{landscape}
\begin{longtable}{llrrrrrl}
\caption{Supernovae used in this work, including their
CMB-frame redshift $z_{\mathrm{CMB}}$,
SALT2 fit parameters $x_1$ and $c$, Hubble residuals
$\Delta\mu_B$, and flux ratios $\mathcal{R}^{(c)}_{642/443}$.
\label{tab:data}
} \\
Name
    & \multicolumn{1}{c}{$z_{\mathrm{CMB}}$}
    & \multicolumn{1}{c}{$x_1$}
    & \multicolumn{1}{c}{$c$}
    & \multicolumn{1}{c}{$\Delta\mu_B$}
    & \multicolumn{1}{c}{$\mathcal{R}_{642/443}$}
    & \multicolumn{1}{c}{$\mathcal{R}^c_{642/443}$}
    & Discovery \\
\hline
\endfirsthead
\caption{continued.} \\
Name
    & \multicolumn{1}{c}{$z_{\mathrm{CMB}}$}
    & \multicolumn{1}{c}{$x_1$}
    & \multicolumn{1}{c}{$c$}
    & \multicolumn{1}{c}{$\Delta\mu_B$}
    & \multicolumn{1}{c}{$\mathcal{R}_{642/443}$}
    & \multicolumn{1}{c}{$\mathcal{R}^c_{642/443}$}
    & Discovery \\
\hline
\endhead
\endfoot

\em{Training SNe} \\
SN2005hc & 0.0449 & $1.21 \pm 0.66$ & $0.016 \pm 0.029$ & $0.113 \pm 0.058$ & $0.393 \pm 0.012$ & $0.385 \pm 0.012$ & SDSS-II; CBET 259 \\
SN2006dm & 0.0208$^{a}$ & $-2.79 \pm 0.38$ & $0.037 \pm 0.032$ & $0.522 \pm 0.111$ & $0.533 \pm 0.007$ & $0.511 \pm 0.007$ & LOSS; CBET 568 \\
SN2007bd & 0.0320$^{b}$ & $-2.90 \pm 0.46$ & $-0.066 \pm 0.026$ & $0.066 \pm 0.073$ & $0.487 \pm 0.010$ & $0.526 \pm 0.011$ & LOSS; CBET 914 \\
SN2007kk & 0.0406$^{c}$ & $-0.20 \pm 0.29$ & $0.058 \pm 0.021$ & $-0.046 \pm 0.060$ & $0.396 \pm 0.007$ & $0.370 \pm 0.006$ & LOSS; CBET 1096 \\
SNF20060908-004 & 0.0490 & $0.60 \pm 0.28$ & $0.047 \pm 0.022$ & $-0.121 \pm 0.229$ & $0.422 \pm 0.010$ & $0.400 \pm 0.010$ \\
SNF20061021-003 & 0.0615 & $-0.34 \pm 0.20$ & $0.088 \pm 0.017$ & $0.198 \pm 0.042$ & $0.489 \pm 0.024$ & $0.441 \pm 0.022$ \\
SNF20061111-002 & 0.0681 & $0.81 \pm 0.36$ & $0.121 \pm 0.020$ & $0.216 \pm 0.039$ & $0.479 \pm 0.023$ & $0.416 \pm 0.020$ \\
SNF20070424-003 & 0.0678 & $0.44 \pm 0.18$ & $0.045 \pm 0.018$ & $0.374 \pm 0.039$ & $0.465 \pm 0.040$ & $0.442 \pm 0.038$ \\
SNF20070531-011 & 0.0365 & $-2.77 \pm 0.39$ & $-0.047 \pm 0.025$ & $0.207 \pm 0.068$ & $0.495 \pm 0.016$ & $0.523 \pm 0.017$ \\
SNF20070630-006 & 0.0708 & $-0.13 \pm 0.17$ & $0.002 \pm 0.018$ & $0.057 \pm 0.039$ & $0.425 \pm 0.036$ & $0.424 \pm 0.036$ \\
SNF20070712-003 & 0.0770 & $0.16 \pm 0.27$ & $-0.050 \pm 0.018$ & $-0.125 \pm 0.146$ & $0.457 \pm 0.027$ & $0.485 \pm 0.029$ \\
SNF20070717-003 & 0.0858 & $-1.09 \pm 0.30$ & $0.060 \pm 0.022$ & $0.373 \pm 0.038$ & $0.586 \pm 0.056$ & $0.546 \pm 0.052$ \\
SNF20070727-016 & 0.0662 & $0.13 \pm 0.22$ & $-0.066 \pm 0.019$ & $-0.376 \pm 0.048$ & $0.349 \pm 0.015$ & $0.377 \pm 0.016$ \\
SNF20070803-005 & 0.0302 & $0.75 \pm 0.28$ & $0.060 \pm 0.027$ & $-0.221 \pm 0.080$ & $0.354 \pm 0.006$ & $0.330 \pm 0.006$ \\
SNF20070806-026 & 0.0439 & $-2.48 \pm 0.22$ & $-0.053 \pm 0.018$ & $0.272 \pm 0.057$ & $0.531 \pm 0.014$ & $0.565 \pm 0.015$ \\
SNF20070810-004 & 0.0823 & $-0.50 \pm 0.20$ & $0.005 \pm 0.017$ & $0.125 \pm 0.034$ & $0.460 \pm 0.023$ & $0.457 \pm 0.023$ \\
SNF20070818-001 & 0.0742 & $-0.68 \pm 0.58$ & $0.096 \pm 0.027$ & $0.201 \pm 0.044$ & $0.499 \pm 0.031$ & $0.445 \pm 0.028$ \\
SNF20071015-000 & 0.0373 & $0.76 \pm 0.31$ & $0.521 \pm 0.025$ & $1.470 \pm 0.067$ & $0.804 \pm 0.035$ & $0.437 \pm 0.019$ \\
SNF20071021-000 & 0.0262 & $-0.90 \pm 0.22$ & $0.126 \pm 0.026$ & $0.471 \pm 0.088$ & $0.572 \pm 0.011$ & $0.494 \pm 0.009$ \\
SNF20080510-001 & 0.0726 & $-0.44 \pm 0.26$ & $-0.032 \pm 0.019$ & $0.043 \pm 0.154$ & $0.481 \pm 0.035$ & $0.499 \pm 0.037$ \\
SNF20080512-010 & 0.0635 & $-3.53 \pm 0.42$ & $-0.036 \pm 0.023$ & $0.221 \pm 0.043$ & $0.490 \pm 0.029$ & $0.512 \pm 0.030$ \\
SNF20080514-002 & 0.0231$^{d}$ & $-2.04 \pm 0.16$ & $-0.079 \pm 0.018$ & $0.234 \pm 0.096$ & $0.478 \pm 0.007$ & $0.524 \pm 0.007$ \\
SNF20080516-022 & 0.0747$^{d}$ & $-0.26 \pm 0.16$ & $-0.058 \pm 0.015$ & $-0.008 \pm 0.035$ & $0.443 \pm 0.048$ & $0.474 \pm 0.051$ \\
SNF20080522-011 & 0.0385 & $1.54 \pm 0.27$ & $0.045 \pm 0.019$ & $-0.152 \pm 0.061$ & $0.456 \pm 0.010$ & $0.433 \pm 0.010$ \\
SNF20080626-002 & 0.0232 & $0.61 \pm 0.31$ & $-0.012 \pm 0.025$ & $-0.048 \pm 0.099$ & $0.426 \pm 0.005$ & $0.432 \pm 0.005$ \\
SNF20080714-008 & 0.0767 & $-0.43 \pm 0.23$ & $0.186 \pm 0.021$ & $0.833 \pm 0.147$ & $0.634 \pm 0.063$ & $0.510 \pm 0.050$ \\
SNF20080803-000 & 0.0568 & $0.28 \pm 0.36$ & $0.147 \pm 0.028$ & $0.520 \pm 0.053$ & $0.573 \pm 0.035$ & $0.482 \pm 0.030$ \\
SNF20080822-005 & 0.0704 & $0.58 \pm 1.07$ & $0.003 \pm 0.029$ & $-0.093 \pm 0.058$ & $0.368 \pm 0.016$ & $0.367 \pm 0.016$ \\
\hline
\em{Validation SNe} \\
SN2004ef & 0.0298$^{e}$ & $-2.40 \pm 0.24$ & $0.152 \pm 0.021$ & $0.594 \pm 0.078$ & $0.645 \pm 0.016$ & $0.540 \pm 0.013$ & Boles, Armstrong; IAUC 8399 \\
SN2005M & 0.0230$^{f}$ & $0.74 \pm 0.34$ & $0.124 \pm 0.026$ & $0.260 \pm 0.101$ & $0.470 \pm 0.008$ & $0.407 \pm 0.007$ & Puckett \& George; IAUC 8470 \\
SN2005di & 0.0242$^{g}$ & $-1.01 \pm 0.25$ & $0.503 \pm 0.037$ & $1.703 \pm 0.102$ & $0.918 \pm 0.041$ & $0.509 \pm 0.023$ & LOSS; CBET 0198 \\
SN2007cq & 0.0245 & $-0.89 \pm 0.25$ & $-0.021 \pm 0.027$ & $-0.135 \pm 0.095$ & $0.406 \pm 0.006$ & $0.416 \pm 0.006$ & Orff \& Newton; CBET 983 \\
SN2007qe & 0.0227 & $-0.36 \pm 0.28$ & $0.119 \pm 0.029$ & $0.302 \pm 0.102$ & $0.462 \pm 0.008$ & $0.402 \pm 0.007$ & ROTSE; CBET 1138 \\
SN2008ec & 0.0151$^{h}$ & $-1.34 \pm 0.22$ & $0.216 \pm 0.028$ & $0.708 \pm 0.149$ & $0.636 \pm 0.006$ & $0.494 \pm 0.005$ & LOSS; CBET 1437 \\
SNF20060511-014 & 0.0467 & $-1.48 \pm 0.23$ & $-0.002 \pm 0.029$ & $0.055 \pm 0.068$ & $0.496 \pm 0.023$ & $0.497 \pm 0.023$ \\
SNF20060526-003 & 0.0787 & $-0.19 \pm 0.56$ & $0.031 \pm 0.029$ & $0.139 \pm 0.042$ & $0.461 \pm 0.025$ & $0.445 \pm 0.024$ \\
SNF20060621-015 & 0.0543 & $0.07 \pm 0.34$ & $-0.065 \pm 0.021$ & $-0.187 \pm 0.047$ & $0.422 \pm 0.012$ & $0.455 \pm 0.013$ \\
SNF20060912-000 & 0.0722 & $-0.13 \pm 0.91$ & $0.141 \pm 0.027$ & $0.191 \pm 0.053$ & $0.498 \pm 0.023$ & $0.423 \pm 0.020$ \\
SNF20060919-007 & 0.0694$^{i}$ & $-0.38 \pm 0.29$ & $-0.008 \pm 0.020$ & $0.015 \pm 0.038$ & $0.459 \pm 0.021$ & $0.463 \pm 0.022$ \\
SNF20061020-000 & 0.0379 & $-2.28 \pm 0.23$ & $0.056 \pm 0.024$ & $0.385 \pm 0.064$ & $0.556 \pm 0.017$ & $0.520 \pm 0.016$ \\
SNF20070403-001 & 0.0815$^{d}$ & $-1.90 \pm 0.30$ & $0.035 \pm 0.016$ & $0.191 \pm 0.034$ & $0.496 \pm 0.042$ & $0.476 \pm 0.041$ \\
SNF20070427-001 & 0.0778$^{d}$ & $0.46 \pm 0.31$ & $-0.115 \pm 0.021$ & $-0.088 \pm 0.036$ & $0.421 \pm 0.025$ & $0.482 \pm 0.029$ \\
SNF20070506-006 & 0.0355 & $0.49 \pm 0.15$ & $-0.017 \pm 0.018$ & $-0.329 \pm 0.065$ & $0.413 \pm 0.011$ & $0.421 \pm 0.011$ \\
SNF20070701-005 & 0.0682 & $-0.88 \pm 0.41$ & $0.039 \pm 0.030$ & $-0.069 \pm 0.050$ & $0.405 \pm 0.024$ & $0.387 \pm 0.023$ \\
SNF20070725-001 & 0.0668 & $0.45 \pm 0.21$ & $-0.105 \pm 0.022$ & $-0.266 \pm 0.050$ & $0.461 \pm 0.021$ & $0.521 \pm 0.024$ \\
SNF20070802-000 & 0.0646 & $0.16 \pm 0.23$ & $0.125 \pm 0.017$ & $0.366 \pm 0.040$ & $0.575 \pm 0.043$ & $0.496 \pm 0.037$ \\
SNF20070817-003 & 0.0630 & $-1.20 \pm 0.24$ & $-0.009 \pm 0.018$ & $0.321 \pm 0.177$ & $0.547 \pm 0.030$ & $0.553 \pm 0.030$ \\
SNF20080323-009 & 0.0834$^{d}$ & $-0.73 \pm 0.23$ & $-0.028 \pm 0.024$ & $0.033 \pm 0.037$ & $0.429 \pm 0.031$ & $0.443 \pm 0.032$ \\
SNF20080510-005 & 0.0850 & $2.29 \pm 0.41$ & $0.123 \pm 0.021$ & $0.191 \pm 0.133$ & $0.455 \pm 0.028$ & $0.395 \pm 0.025$ \\
SNF20080516-000 & 0.0726 & $0.88 \pm 0.24$ & $-0.007 \pm 0.019$ & $-0.076 \pm 0.155$ & $0.415 \pm 0.025$ & $0.419 \pm 0.025$ \\
SNF20080522-000 & 0.0460 & $0.96 \pm 0.20$ & $0.076 \pm 0.019$ & $-0.225 \pm 0.242$ & $0.363 \pm 0.012$ & $0.332 \pm 0.011$ \\
SNF20080531-000 & 0.0369 & $-0.08 \pm 0.33$ & $0.037 \pm 0.022$ & $0.282 \pm 0.064$ & $0.504 \pm 0.014$ & $0.482 \pm 0.014$ \\
SNF20080610-000 & 0.0789$^{d}$ & $-0.48 \pm 0.23$ & $-0.005 \pm 0.019$ & $0.150 \pm 0.038$ & $0.477 \pm 0.029$ & $0.480 \pm 0.030$ \\
SNF20080612-003 & 0.0329 & $0.15 \pm 0.20$ & $-0.011 \pm 0.022$ & $-0.284 \pm 0.072$ & $0.450 \pm 0.011$ & $0.456 \pm 0.011$ \\
SNF20080614-010 & 0.0752$^{d}$ & $-2.89 \pm 0.18$ & $-0.002 \pm 0.018$ & $0.340 \pm 0.036$ & $0.488 \pm 0.037$ & $0.489 \pm 0.037$ \\
SNF20080623-001 & 0.0427 & $-0.70 \pm 0.18$ & $-0.050 \pm 0.017$ & $0.262 \pm 0.260$ & $0.483 \pm 0.013$ & $0.512 \pm 0.014$ \\
SNF20080720-001 & 0.0200 & $0.30 \pm 0.26$ & $0.650 \pm 0.025$ & $1.436 \pm 0.156$ & $0.861 \pm 0.015$ & $0.402 \pm 0.007$ \\
SNF20080810-001 & 0.0426 & $-1.37 \pm 0.26$ & $0.038 \pm 0.026$ & $0.274 \pm 0.060$ & $0.529 \pm 0.016$ & $0.506 \pm 0.015$ \\
\hline
\em{Literature SNe} \\
SN1998V & $0.0172$  & \multicolumn{1}{c}{...} & \multicolumn{1}{c}{...}  & $0.412 \pm 0.167$  & $0.423 \pm 0.017$  & \multicolumn{1}{c}{...}  & Armstrong; IAUC 6841 \\
SN1998aq & $0.0045$  & \multicolumn{1}{c}{...} & \multicolumn{1}{c}{...}  & $0.002 \pm 0.558$  & $0.404 \pm 0.013$  & \multicolumn{1}{c}{...}  & Armstrong; IAUC 6875 \\
SN1998bp & $0.0102$  & \multicolumn{1}{c}{...} & \multicolumn{1}{c}{...}  & $1.210 \pm 0.217$  & $0.723 \pm 0.106$  & \multicolumn{1}{c}{...}  & Armstrong; IAUC 6890 \\
SN1998de & $0.0156$  & \multicolumn{1}{c}{...} & \multicolumn{1}{c}{...}  & $2.303 \pm 0.141$  & $0.992 \pm 0.239$  & \multicolumn{1}{c}{...}  & LOSS; IAUC 6977 \\
SN1998dh & $0.0077$  & \multicolumn{1}{c}{...} & \multicolumn{1}{c}{...}  & $0.515 \pm 0.270$  & $0.553 \pm 0.190$  & \multicolumn{1}{c}{...}  & LOSS; IAUC 6978 \\
SN1998ec & $0.0201$  & \multicolumn{1}{c}{...} & \multicolumn{1}{c}{...}  & $0.628 \pm 0.156$  & $0.667 \pm 0.024$  & \multicolumn{1}{c}{...}  & BAO SN Survey; IAUC 7022 \\
SN1998eg & $0.0235$  & \multicolumn{1}{c}{...} & \multicolumn{1}{c}{...}  & $0.467 \pm 0.141$  & $0.550 \pm 0.190$  & \multicolumn{1}{c}{...}  & Boles; IAUC 7033 \\
SN1998es & $0.0096$  & \multicolumn{1}{c}{...} & \multicolumn{1}{c}{...}  & $-0.108 \pm 0.212$  & $0.452 \pm 0.012$  & \multicolumn{1}{c}{...}  & LOSS; IAUC 7050 \\
SN1999aa & $0.0153$  & \multicolumn{1}{c}{...} & \multicolumn{1}{c}{...}  & $-0.231 \pm 0.155$  & $0.356 \pm 0.023$  & \multicolumn{1}{c}{...}  & Arbour; IAUC 7108 \\
SN1999ac & $0.0098$  & \multicolumn{1}{c}{...} & \multicolumn{1}{c}{...}  & $0.096 \pm 0.235$  & $0.419 \pm 0.004$  & \multicolumn{1}{c}{...}  & LOSS; IAUC 7144 \\
SN1999cc & $0.0315$  & \multicolumn{1}{c}{...} & \multicolumn{1}{c}{...}  & $0.232 \pm 0.073$  & $0.519 \pm 0.033$  & \multicolumn{1}{c}{...}  & Schwartz; IAUC 7163 \\
SN1999cl & $0.0087$  & \multicolumn{1}{c}{...} & \multicolumn{1}{c}{...}  & $1.141 \pm 0.292$  & $1.890 \pm 0.245$  & \multicolumn{1}{c}{...}  & LOSS; IAUC 7185 \\
SN1999ej & $0.0128$  & \multicolumn{1}{c}{...} & \multicolumn{1}{c}{...}  & $0.823 \pm 0.195$  & $0.569 \pm 0.005$  & \multicolumn{1}{c}{...}  & LOSS; IAUC 7286 \\
SN1999gd & $0.0193$  & \multicolumn{1}{c}{...} & \multicolumn{1}{c}{...}  & $1.302 \pm 0.162$  & $0.946 \pm 0.190$  & \multicolumn{1}{c}{...}  & LOSS; IAUC 7319 \\
SN2000dk & $0.0164$  & \multicolumn{1}{c}{...} & \multicolumn{1}{c}{...}  & $0.225 \pm 0.136$  & $0.545 \pm 0.126$  & \multicolumn{1}{c}{...}  & LOSS; IAUC 7493 \\
SN2000fa & $0.0218$  & \multicolumn{1}{c}{...} & \multicolumn{1}{c}{...}  & $0.004 \pm 0.151$  & $0.433 \pm 0.093$  & \multicolumn{1}{c}{...}  & LOTOSS; IAUC 7533 \\
\hline
\multicolumn{8}{p{\textwidth}}{
{\it Notes}:
The Hubble residuals
are {\em before} any corrections for $c$, $x_1$, or $\mathcal{R}$.
The peak magnitudes for the literature SNe Hubble diagram fits
are from \cite{HickenLC} and \cite{Jha_LC}, which did not use SALT2.
The $\mathcal{R}_{642/443}$ Hubble diagram fits were performed separately for the
literature and SNfactory observations; the reported $\Delta\mu_B$
residuals are with respect to the individual fits.
The full table available from CDS includes all flux ratios from
Table~\ref{tab:results} for the SNe observed by the SNfactory.
} \\
\multicolumn{8}{p{\textwidth}}{
Discoveries and redshifts are from SNfactory observations
except as otherwise noted.
Literature SNe redshifts are via \cite{HickenLC} and \cite{Jha_LC}.
Other redshift citations via NED except for SDSS: \newline
\hspace*{1 em} $^{a}$Theureau, G. et al.\ 1998, \aaps, 130, 333; \newline
\hspace*{1 em} $^{b}$Falco, E.~E., et al.\ 1999, \pasp, 111, 438; \newline
\hspace*{1 em} $^{c}$Huchra, J.~P., Vogeley, M.~S., \& Geller, M.~J.\ 1999, \apjs, 121, 287; \newline
\hspace*{1 em} $^{d}$SDSS; \newline
\hspace*{1 em} $^{e}$O'Neil, K.\ 2004, \aj, 128, 2080; \newline
\hspace*{1 em} $^{f}$de Vaucouleurs, G. et al.\ 1992, VizieR Online Data Calalog, 7137, 0; \newline
\hspace*{1 em} $^{g}$Mathewson, D.~S., \& Ford, V.~L.\ 1996, \apjs, 107, 97; \newline
\hspace*{1 em} $^{h}$Keel, W.~C.\ 1996, \apjs, 106, 27; \newline
\hspace*{1 em} $^{i}$Jones, D.~H. et al.\ 2005, PASA, 22, 277; \newline
} 

\end{longtable}
\end{landscape}
} 


\begin{thebibliography}{99}
%
\bibitem[Aldering et al.(2002)]{SNF_SPIE}
Aldering, G., Adam, G., Antilogus, P. et al.
2002, \procspie, 4836, 61



\bibitem[Blondin et al.(2009)]{Blondin_99cl}
Blondin, S., Prieto, J.~L., Patat, F., et al. 
2009, \apj, 693, 207

\bibitem[Bongard et al.(2006)]{Bongard_RSiS}
Bongard, S., Baron, E., 
Smadja, G., Branch, D., \& Hauschildt, P.~H.\ 2006, \apj, 647, 513 

\bibitem[Bongard et al.(2008)]{Bongard_PHOENIX}
Bongard, S., Baron, E., 
Smadja, G., Branch, D., \& Hauschildt, P.~H.\ 2008, \apj, 687, 456 

\bibitem[Bronder et al.(2008)]{Bronder_EWSiII}
Bronder, T.~J., Hook, I.~M., Astier, P. et al.
2008, \aap, 477, 717 

\bibitem[Cardelli et al.(1989)]{CCM}
Cardelli, J.~A., Clayton, G.~C., \& Mathis, J.~S.\ 1989, \apj, 345, 245 

\bibitem[Foley et al.(2008)]{Foley_UV}
Foley, R.~J., Filippenko, A.~V., \& Jha, S.~W.\ 2008, \apj, 686, 117 

\bibitem[Guy et al.(2007)]{Guy_SALT2}
Guy, J., Astier, P., Baumont, S. et al.
2007, \aap, 466, 11 

\bibitem[Hachinger et al.(2006)]{Hachinger_EW}
Hachinger, S., Mazzali, P.~A., \& Benetti, S.\ 2006, \mnras, 370, 299 

\bibitem[Hicken et al.(2009)]{HickenLC}
Hicken, M., Challis, P.,  Jha, S. et al.
2009, arXiv:0901.4787 

\bibitem[Jha et al.(2006)]{Jha_LC}
Jha, S., Kirshner, R.~P., Challis, P. et al.
2006, \aj, 131, 527 

\bibitem[Jha et al.(2007)]{Jha_MLCS}
Jha, S., Riess, A.~G., \& Kirshner, R.~P.\ 2007, \apj, 659, 122 

\bibitem[Kasen et al.(2006)]{Kasen_SEDONA}
Kasen, D., Thomas, R.~C., \& Nugent, P.\ 2006, \apj, 651, 366 

\bibitem[Kowalski et al.(2008)]{KowalskiUnion2008} Kowalski, M., et al.\ 
2008, \apj, 686, 749 

\bibitem[Krisciunas et al.(2006)]{Krisciunas_99cl_dust}
Krisciunas, K.,  Prieto, J.~L., Garnavich, P.~M., et al.
2006, \aj, 131, 1639 

\bibitem[Matheson et al.(2008)]{MathesonSpec}
Matheson, T., Kirshner, R.~P., Challis, P. et al.
2008, \aj, 135, 1598 

\bibitem[Nugent et al.(1995)]{Nugent_RSi}
Nugent, P., Phillips,  M., Baron, E., Branch, D., \& Hauschildt, P.\ 1995,
\apjl, 455, L147 

\bibitem[Phillips(1993)]{Phillips93}
Phillips, M.~M.\ 1993, \apjl, 413, L105 

\bibitem[Prieto et al.(2006)]{Prieto2006} Prieto, J.~L., Rest, A., 
\& Suntzeff, N.~B.\ 2006, \apj, 647, 501 

\bibitem[Riess et al.(1998)]{SnapshotDistances}
Riess, A.~G., Nugent, P.,  Filippenko, A.~V., Kirshner, R.~P.,
\& Perlmutter, S.\ 1998, \apj, 504, 935 

\bibitem[Riess \& Livio(2006)]{RiessLivio2006}
Riess, A.~G., \& Livio, M.\ 2006, \apj, 648, 884 

\bibitem[Schlegel et al.(1998)]{SFD}
Schlegel, D.~J., Finkbeiner, D.~P., \& Davis, M.\ 1998, \apj, 500, 525 

\end{thebibliography}
\end{document}